\input harvmac.tex
\noblackbox

\lref\LSone{F. Lesage, H. Saleur, ``Correlations in
one dimensional quantum impurity problems with an external field'',
submitted to Nucl. Phys. B; cond-mat/9611025.}
\lref\korbook{V.E. Korepin,
N.M. Bogoliubov, A.G. Izergin, ``Quantum inverse scattering
method and correlation functions", Cambridge University Press,
(1993).}
\lref\smirnov{F.A. Smirnov, ``Form factors in completely integrable
models of quantum field theory", Worlk Scientific, and references
therein.}
\lref\flsbig{P. Fendley, A.W.W. Ludwig, H. Saleur, Phys. Rev.
B52, 8934 (1995), cond-mat/9503172.}
\lref\LSS{F. Lesage, H. Saleur, S. Skorik, Phys.
Rev. Lett. 76 (1996) 3388, cond-mat/9512087 ;  Nucl.
Phys. B474 (1996), 602,
cond-mat/9603043 .}
\lref\LLSS{A. Leclair, F. Lesage, S. Sachdev, H. Saleur, to appear in
Nucl. Phys. B., cond-mat/9606104.}
\lref\DMSmassless{G. Delfino, G. Mussardo, P. Simonetti, Phys. Rev.
D51, 6620 (1995).}
\lref\FSnoise{P. Fendley,
H. Saleur,
to appear in
Phys. Rev. B., cond-mat/9601117.}
\lref\subir{S. Sachdev, K. Damle, ``Low temperature spin diffusion in
the one-dimensional quantum $O(3)$ nonlinear $\sigma$ model'',
cond-mat/9610115.}
\lref\ZZ{A.B. Zamolodchikov and Al.B. Zamolodchikov, Ann. Phys. 120
(1979) 253.}
\lref\LB{ R. Landauer, Physica D38 (1989) 594;
G.B. Lesovik, JETP Lett. 49 (1989) 594; M. B\"uttiker,
Phys. Rev. Lett. 65 (1990) 2901.}

\Title{USC-96-028}
{\vbox{
\centerline{Correlations in one dimensional quantum impurity
}
 \vskip 4pt
\centerline{problems with an external field or a temperature.}
}}

\centerline{F. Lesage, H. Saleur.}
\bigskip\centerline{Department of Physics,}
\centerline{University of Southern California,}
\centerline{Los Angeles, CA 90089-0484}
\centerline{PACS: 71.10.Pm, 71.15.Cr, 72.10.Fk.}

\vskip .3in
We  discuss in more details the theory of low energy
excitations in  quantum impurity problems with an external field
at vanishing temperature, giving further support
to results of the previous paper. We then extend
these results to the next order  at low frequency,
obtaining in particular the exact expression, as a function of the
bias, of the
first two derivatives of the response function $\chi''(\omega)$
at $\omega=0$ in the double well problem of dissipative quantum
mechanics.

We also extend our approach to the case of non vanishing
temperature and no external field.
Fendley et al. had obtained in that case
an expression for
the Hall conductance with a single impurity using a
Landauer-B\"uttiker
type approach. We recover their result in the framework
of linear response theory using renormalized form-factors.
We also obtain, as a function of the temperature, the
first  derivative of the response function $\chi''(\omega)$
at $\omega=0$ in the double well problem of dissipative quantum
mechanics.

\Date{12/96}

\newsec{Introduction}

A lot of progress has been accomplished recently in the
non perturbative  computation
of correlation functions for integrable quantum field theories
in 1+1 dimensions \smirnov,\korbook,\DMSmassless. Interesting
physical  applications
have been found in particular in one dimensional quantum impurity
problems \ref\guinea{F. Guinea, V. Hakim, A. Muramatsu, 
Phys. Rev. B32, 4410 (1985); S.A. Bulgadaev, JEPT vol. 38, 264 (1984),
Sov. Phys. JETP, vol. 39, 314 (1984).}, \flsbig,\LSS.
Unfortunately, many of these computations
have been restricted to the case of vanishing temperature
and external field, preventing in most cases
comparison with experimental data; a notable exception
concerns the DC properties for tunneling between  edge states
in fractional quantum Hall devices \flsbig.

The introduction of a temperature or an external field, while
it does not break integrability, renders the computation
of correlators using form-factors  more difficult (putting a
temperature
seems somewhat
more natural in the approach of \korbook, which however produces only
rather implicit results).   Few works have
addressed this question so far,
except in the free fermion case,
where however the problem is already quite non trivial for
spin operators \LLSS.  We note that
some   interesting results have also been obtained
in the case of interacting massive bulk theories \subir\ in the limit
of very low temperatures. In \LSone\ (henceforth refered to as I)
we have considered quantum impurity problems at $T=0$
but with an external field.
We were able to
determine two important low energy
properties, the limit $\lim_{\omega\to 0}{\chi''(\omega)\over
\omega}$
of the dynamical susceptibiliy in the double well problem of
dissipative
quantum mechanics, and the $|\omega|$ component of the noises in a
four
terminal geometry for  tunneling between edges in  fractional
quantum Hall devices.
Some global properties, like the existence of potential
singularities,
were also examined.

The first purpose of this sequel is to place
the low energy analysis of I on firmer grounds, and to compute
the next order at low frequency, eg the $\omega^3$ term
in $\chi''(\omega)$.   This is the subject of sections 2 and 3.

Our second purpose is to carry out a similar analysis for the case of
a non vanishing temperature (section 4). We discuss the formula for
the
conductance through a point contact that
had been obtained previously by Fendley et al.
using a Landauer-B\"uttiker \LB\  type approach,
and recover it, at the price of some reasonable hypothesis, in
the framework of linear response theory. Building on this analysis,
we obtain the limit $\lim_{\omega\to 0}{\chi''(\omega)\over \omega}$
of the dynamical susceptibiliy in the double well problem of
dissipative quantum mechanics at finite temperature.

\newsec{A closer look at low energy excitations with $T=0,V\neq 0$}

This paper is a sequel to I and the notation introduced there
will be used here. The hamiltonian we study,
\eqn\hamil{
H={1\over 2}\int_{-\infty}^0 dx \ [8\pi g\Pi^2+{1\over 8\pi g}
(\partial_x\phi)^2]+H_B,
}
 consists of a free bulk theory with a boundary
interaction.
The detailed  form of this interaction, $H_B$, is
different for the  dissipative quantum mechanics and the
tunneling  problems.

The principle of our approach using
massless scattering is described in full details in I.
In brief, the bulk degres of freedom are described by a massless
sine-Gordon theory which has solitons/anti-solitons and
bound states as fundamental  ``bare'' excitations.  In the presence
of
a voltage, the ground state
consists of a Fermi sea filled with solitons, in a way
easily controlled using the Bethe ansatz. Physical properties
of interest  are  determined by the structure of excitations
over this voltage dependent ground state. The boundary interaction
is finally  taken into account by the introduction of reflection
matrices.

In this section
we discuss the voltage dependent  ground state and its excitations
more thoroughly. The main problem we are  addressing is the
scattering of these excitations, which is ``dressed'' by the presence
of
the Fermi sea, and the resulting dressed form-factors.
Most of the discussion is centered on the scattering at coincident
rapidities, where, for non-neutral excitations,
a  non trivial phase seems to  appear.
This phase, and its possible effects on the  low energy properties,
are discussed in details.

\subsec{The Fermi sea}

Recall that at zero temperature and in the presence of
an external field, the ground state contains only solitons.
The model is massless and a convenient way to use the
Bethe ansatz in this problem is to ``unfold" the
half-line problem to obtain a  problem on the full line.  Then we
can, for instance, consider a system with only
right movers.
The system is therefore described
by the Bethe ansatz equations (setting $\hbar=1$)
\eqn\bethedeux{ I_\alpha= {\mu L\over 2\pi}
e^{\theta_\alpha}+\sum_\beta
\delta(\theta_\alpha-\theta_\beta),}
where we have reinstated an arbitrary energy scale $\mu$ (set equal
to one in I) and $\delta$ is the shift coming from the
soliton-soliton S matrix.
In the ground state, solitons fill the interval
$\theta\in [-\infty,A]$\foot{As in I we use rapidities to parametrize
the energy and momentum of a particle.},
with a density $\rho$  obeying
\eqn\denI{\rho(\theta)={\mu e^\theta\over 2\pi}+
\int_{-\infty}^A \Phi(\theta-\theta')\rho(\theta')
d\theta',}
where $S_{++}^{++}=S=e^{2i\pi\delta},\ \Phi(\theta)={1\over 2i\pi}
{d\ln S\over d\theta}$.
It should be pointed out that the quantization condition
\bethedeux\ should also involve the reflection matrix but
its contribution to the density is of order $1/L$, and this
is negligible for our purposes.
Equations of the type \denI\ for a general function $f$
$$
f(\theta)=g(\theta)+\int_{-\infty}^A
\Phi(\theta-\theta')f(\theta')d\theta'
$$
will occur repeatedly in the subsequent analysis.
Let us introduce the functional $\hat{K}$ such that
$$
\hat{K}\bullet f(\theta)=\int_{-\infty}^A
\Phi(\theta-\theta')f(\theta')d\theta'
$$
and similarly
$$
\hat{I}\bullet f(\theta)=f(\theta)
$$
Then the previous equation reads $(\hat{I}-\hat{K})\bullet f=g$,
and the solution follows by introducing an operator $L$ such that
$$
(\hat{I}+\hat{L})(\hat{I}-\hat{K})=\hat{I},
$$
that is
$$
f(\theta)=g(\theta)+\int_{-\infty}^A L(\theta,\theta')
g(\theta')d\theta'.
$$
The functions $L$ is symmetric of its
two arguments $L(\theta,\theta')=L(\theta',\theta)$ and can be
written formally as
\eqn\formal{
L(\theta,\theta')=\Phi(\theta-\theta')+\int_{-\infty}^A
\Phi(\theta-\theta'')
\Phi(\theta''-\theta')d\theta''+\ldots.}
It is important to stress that $L$ is not a function of the
difference
of its arguments, as the previous equation clearly shows.
Observe that one has  $(\hat{I}+\hat{L})\hat{K}=\hat{L}$,
or more explicitly
$$
\int_{-\infty}^A
[\delta(\theta-\theta')+L(\theta,\theta')]
\Phi(\theta'-\theta'')d\theta'=
L(\theta,\theta'').
$$

In the particular case of the density we have $\rho(\theta)=
{\mu\over 2\pi}
\left[e^\theta+\int_{-\infty}^A
 L(\theta,\theta') \ e^{\theta'}d\theta'\right]$. For $\theta<A$,
$\rho$ is the density of solitons in the sea.
 For $\theta>A$, $\rho$ is the density
of holes of  solitons above the sea. The function
$\rho$ can be written maybe more explicitely as an infinite series
of exponentials using Wiener Hopf integration
techniques, see I. We recall  the value
$\rho(A)={V\over 4\pi}\sqrt{2g}$.
The Fermi rapidity $A$ is determined self consistently by introducing
the quantity
\eqn\epsI{\epsilon(\theta)=\mu e^\theta-{V\over 2}+
\int_{-\infty}^A \Phi(\theta-\theta')\epsilon(\theta')d\theta',}
and requiring $\epsilon(A)=0$. Observe the identity
\eqn\idenI{{d\epsilon(\theta)\over d\theta}=2\pi \rho(\theta).}

\subsec{Neutral  excitations}

Let us now discuss excitations over the ground state.
First, consider
a particle hole excitation. In order not to make
confusion with the other types of particles (eg antisolitons)
which also appear in this problem at high energy,  we call a soliton
 above the
sea a {\bf volton}, and a hole in the sea an {\bf antivolton}.
Suppose the volton has  rapidity $\theta_p$ and the antivolton
rapidity
$\theta_h$. The particles are interacting
and this induces a shift of the rapidities
in the sea: a rapidity equal to $\theta_\alpha$ initially becomes
$\theta_\alpha
+\delta^{(2)}\theta_\alpha$ with the conditions
\eqn\conds{\eqalign{I_\alpha=&{\mu L\over 2\pi}e^{\theta_\alpha}+
\sum_\beta \delta(\theta_\alpha-\theta_\beta)\cr
I_\alpha=&{\mu L\over
2\pi}e^{\theta_\alpha+\delta^{(2)}\theta_\alpha}+
\sum_\beta
\delta(\theta_\alpha-\theta_\beta+\delta^{(2)}
\theta_\alpha-\delta^{(2)}
\theta_\beta) \cr
&+\delta(\theta_\alpha-\theta_p)-
\delta(\theta_\alpha-\theta_h).\cr}}
We then define the shift function, describing the change in the
Fermi sea, by
$L\rho(\theta)\delta^{(2)}\theta\equiv F(\theta|\theta_p,
\theta_h)$.
By standard manipulations \korbook, one finds
the equation obeyed by the shift
\eqn\shifti{F(\theta|\theta_p,\theta_h)-
\int_{-\infty}^A
\Phi(\theta-\theta')F(\theta'|\theta_p,\theta_h)d\theta'
=\delta(\theta-\theta_h)-
\delta(\theta-\theta_p).}
A formal solution of this equation
follows as
\eqn\form{F(\theta|\theta_p,\theta_h)=
\int_{\theta_h}^{\theta_p}L(\theta,\theta')d\theta'.}
The change of energy of the system with this excitation
is given by the bare energies of the particle and the hole
and the contribution from the shift of the Fermi sea
\eqn\trivI{\delta^{(2)}E=\mu e^{\theta_p}-\mu
e^{\theta_h}+\int_{-\infty}^A F(\theta|
\theta_p,\theta_h) \ \mu e^\theta d\theta,}
and this can be shown  \korbook\ to coincide with
\eqn\trivII{\delta^{(2)}E=\epsilon(\theta_p)-\epsilon(\theta_h),}
with $\epsilon$ defined in \epsI .
One has  $\epsilon(\theta)<0$ for $\theta<A$, so the energy of hole
excitations
is positive as it should.  We note the value $\epsilon(-\infty)=-\mu
gV$.
Similar computations can be carried out for the momentum, with the
result that, for these neutral excitations,
$\delta^{(2)}P=\delta^{(2)}E$; hence the excitations are massless,
as physically expected.

\subsec{Non neutral excitations}

Although physical excitations will always be made of
volton-antivolton
pairs, it is necessary, to understand the scattering properties,
to consider excitations with only one volton or one antivolton.
Such excitations will involve a ``Fermi'' momentum, that is
$\delta^{(1)}P=\delta^{(1)}E\pm p_f$. Assuming for the moment
that the excitation energy of a volton (resp. an antivolton) is still
given by $\epsilon$ (resp. -$\epsilon$), let us now determine the
Fermi momentum.  Recall that the proper definition of $p$ is via the
phase shift collected when a particle goes around the world.
For a volton excitation this phase reads
$$
Lp=L\mu e^\theta+2\pi\sum_\alpha\delta(\theta-\theta_\alpha^{(1)})+
2\pi\sum_\alpha \delta(\infty).
$$
Here, we have  taken the usual phase shift $\delta={1\over 2i\pi}\ln
S$
where $S$ is the soliton soliton sine-Gordon S-matrix. However, when
taking the massless limit of the sine-Gordon model,
left and right movers do not become totally independent.
There remains a RL constant scattering
phase, $\delta(\infty)$.
This phase appears very rarely
in computations, and its meaning is not totally clear. But when we
pass a particle through the system, this RL shift
contributes to the momentum,
and this is the meaning of the last term in the
foregoing equation.  At leading order, we can neglect the shift of
rapidities in the sea and reexpress this as
$$
Lp=L\mu
e^\theta+2\pi\int_{-\infty}^A[\delta(\theta-\theta')+
\delta(\infty)]\rho(\theta')d\theta'
$$
Then, using the relation
$\rho={1\over 2\pi}{d\epsilon\over d\theta}$, we can write
$$
Lp=L\epsilon(\theta)+L{V\over 2}+2\mu LgV\delta(\infty).
$$
One checks that $\delta(\infty)=-\delta(-\infty)={1\over 2}-{1\over
4g}$, and it
follows that
\eqn\fermimom{p_f=\mu gV.}
Consider now the operator destroying a right moving volton
and creating a left moving one. Its correlation function,
from this extra Fermi momentum,
will be alternating with a $cos 2p_Fx$ part. On the other
hand,
the alternating part can be computed by observing that the potential
$V$ can be absorbed by a
redefinition of the field $\phi_{L,R}\to \phi_{L,R}\pm gVx$ in the
original hamiltonian.
This leads to the
identification of this operator with $\exp i(\phi_L-\phi_R)$, of
conformal weights
$h=\bar{h}=g$. More generally, the operator $\exp
i\alpha(\phi_L-\phi_R)$
has conformal weight $h=g\alpha^2$ and its correlation
function involves a part $\cos 2\alpha gVx$. This corresponds
to a charge $Q=2g\alpha$, in conventions where the soliton has unit
charge .

Notice that the LR phase shift should appear in \bethedeux. However,
as long as the number of L particles is a constant, its presence
would simply shift all the $I_\alpha$ by a constant, without
changing the results for  densities or excitation energies. Things
are different when the number of particles is changed. For instance,
adding a R soliton adds up a phase $\delta(-\infty)$ to the rhs
of Bethe equations for L movers. Without any correction
term, this phase in turn will move the Fermi sea, resulting
in an induced L charge, an effect which is not
expected for a purely R excitation. This means that, to observe
physical non neutral excitations, one has to complement
the addition or removal of a particle by
an additional phase shift,  corresponding
presumably to changing boundary conditions.
Determining the value of this phase is not obvious and, in
the rest of this section, we
embark in a long discussion of the properties of the system
with this phase to see its effect.
Consider for instance a volton excitation, associated
with an additional phase shift $\exp 2i\pi\delta_b$. The
following holds
\eqn\condsi{\eqalign{I_\alpha=&{\mu L\over 2\pi}e^{\theta_\alpha}+
\sum_\beta \delta(\theta_\alpha-\theta_\beta)\cr
I_\alpha+\delta_b=&{\mu L\over
2\pi}e^{\theta_\alpha+\delta^{(1)}\theta_\alpha}+
\sum_\beta
\delta(\theta_\alpha-\theta_\beta+\delta^{(1)}
\theta_\alpha-\delta^{(1)}
\theta_\beta) \cr
&+\delta(\theta_\alpha-\theta_p),\cr}}
with $\theta_p$ the volton rapidity.
By the same logic as before, defining
$F_b(\theta|\theta_p)=L\rho(\theta)\delta^{(1)}\theta$,
one has
\eqn\shifti{F_b(\theta|\theta_p)-
\int_{-\infty}^A\Phi(\theta-\theta')F_b(\theta'|\theta_p)
=\delta_b-\delta(\theta-\theta_p).}
If we were to add an antivolton, the additional phase shift
would be $\exp -2i\pi\delta_b$, so of course for pairs voltons
antivoltons
the total shift would be as before, as one has
$$
F(\theta|\theta_p,\theta_h)=F_b(\theta|\theta_p)-F_b(\theta|\theta_h)
$$
for any $\delta_b$, so \trivI,\trivII\ hold indenpendently of this
phase.
Now, for non ``neutral'' excitations we have to require more,
namely that the excitation energy also is given by
$\delta^{(1)}E=\epsilon(\theta_p)$.
Let us see whether this can be satisfied:
On the one hand, the solution of \epsI\ reads
\eqn\epsIsol{\epsilon(\theta_p)= \mu e^{\theta_p}
-{V\over 2}+\int_{-\infty}^A \mu e^{\theta}
L(\theta_p,\theta)d\theta-{V\over 2}\int_{-\infty}^A
L(\theta_p,\theta)d\theta.}
On the other hand, the excitation energy due to the volton can be
written
\eqn\dirI{\varepsilon(\theta_p)=\mu e^{\theta_p}-{V\over 2}+
\int_{-\infty}^A \mu e^{\theta'}F_b(\theta'|\theta_p)d\theta'-{V\over
2}F_b(-\infty|\theta_p).}
The last term was not present for neutral excitations.
First, using the formal solution \formal , one checks that
$F_b(-\infty,\theta_p)$ is in fact independent of  $\theta_p$.
For neutral excitations,
this   term would be cancelled by its counterpart
due to the antivolton.  The meaning of $F_b(-\infty,\theta_p)$ is
simple:
it is  (minus) the number of solitons that are pushed ``out'' of the
Fermi sea
in the presence of an added soliton above the sea. While these
particles have
a vanishing kinetic energy, they have a non vanishing potential
energy,
which must be put by hand.

With the shift $\delta_b$, the solution of \shifti\ reads
\eqn\explic{F_b(\theta|\theta_p)=[\delta_b-\delta(\infty)]\left[
1+\int_{-\infty}^A L(\theta,\theta')d\theta'\right]+
\int_{-\infty}^{\theta_p}L(\theta,\theta')d\theta'.}
{}From the defining equation for $F_b$ \shifti, we can also deduce
\eqn\derivv{\eqalign{{d\over d\theta}F_b(\theta|\theta_p)=
&-L(\theta,\theta_p)-L(\theta,A)
F_b(A|\theta_p)\cr
{d\over d\theta_p}F_b(\theta|\theta_p)=&L(\theta,\theta_p).\cr}}
Integration by parts in \dirI\ givves rise to the alternate
expression
\eqn\altdirI{\eqalign{
\varepsilon(\theta_p)=\mu e^{\theta_p}-&{V\over 2}+\mu
e^AF_b(A|\theta_p) \cr
+\int_{-\infty}^A
\mu e^{\theta}L(\theta,\theta_p)d\theta &
+F_b(A|\theta_p)\int_{-\infty}^A\mu e^\theta L(\theta,A)d\theta
-{V\over 2}
F_b(-\infty|\theta_p).}}
Using $\epsilon(A)=0$, the equality $\epsilon=\varepsilon$ will
hold iff
\eqn\mainicond{F_b(A|\theta_p)
\left[1+\int_{-\infty}^AL(A,\theta)d\theta\right]
-F_b(-\infty|\theta_p)=-\int_{-\infty}^A L(\theta_p,\theta)d\theta.}
The left hand side is linear in $\delta_b$ with  the coefficient
$$
\left[1+\int_{-\infty}^A L(A,\theta)d\theta\right]^2-1
-\int_{-\infty}^AL(-\infty,\theta) \ d\theta
$$
(for the second term, the upper integration bound has no importance
since the region where the integrand is finite
is concentrated around $-\infty$).  By using the equations for $\rho$
and
$\epsilon$ together with the values of $\rho(A)$ and
$\epsilon(-\infty)$
one finds the results
\eqn\numeri{
\eqalign{\int_{-\infty}^A L(A,\theta)d\theta=&\sqrt{2g}-1\cr
\int_{-\infty}^A L(-\infty,\theta)d\theta=&2g-1.\cr}}
Hence the left hand side of \mainicond\ actually
is independent of $\delta_b$.
Let us chose the particular value $b^*$ of  $\delta_b$ such that
$F_{b^*}(A,A)=0$.
Using \derivv\ it follows that in that case
$F_{b^*}(\theta,A)=-F_{b^*}
(A,\theta)$. One has
then
\eqn\Fpart{F_{b^*}(\theta_p|A)=
{\int_{-\infty}^A [L(\theta_p,\theta)-L(A,\theta)]d\theta
\over 1+\int_{-\infty}^A L(A,\theta)d\theta},}
from which \mainicond\ follows together with
$F_{b^*}(-\infty,\theta)=\int_{-\infty}^A
L(A,\theta)d\theta=\sqrt{2g}-1$. Hence, indeed, the function
$\epsilon$
gives the excitation energy of non neutral excitations too, and this
for any value of $\delta_b$.

Another quantity of interest is the charge of excitations.
 For a volton for instance,
it is equal to $+1$, the bare charge, minus the number of solitons
ejected from the sea at $-\infty$, which reads
$-F_b(-\infty|\theta_p)$.
 So we have
\eqn\newcharge{q_{\pm}(\theta)=\pm [1+F_b(-\infty|\theta)]}
where
from now on, we designate voltons and antivoltons by a label $\pm$.
The modulus of this charge is a constant, so neutral excitations
still consist of pairs voltons-antivoltons, as physically expected.
For a generic choice of shift $\delta_b$, one finds that
$F_b(-\infty|\theta)=2g[\delta_b-\delta(\infty)]+2g-1$. For this
value, one has
$F_b(A|A)=\sqrt{2g}[\delta_b-\delta(\infty)]+\sqrt{2g}-1$.

At this stage, it is useful to consider the effect of shifting the
edge of the Fermi sea. We would like to compute the
corresponding change in energy to order $1/L$ and see if
that will furnish more constraints on $\delta_b$.
So far, we did all computations
without worrying about such terms. It is for instance not possible to
use straightforwardly the formula for excitation energies $\epsilon$
when
one wants such corrections. The safest is to go back to
original definitions. The energy reads as a discrete sum. When
replacing the sum
by an integral in the ground state, the Euler-Mac Laurin formula
can be applied. It shows that there is a term proportional
to $L$, no term of order one (as physically expected when
there is no boundary nor impurity), and a term of order $1/L$ that
determines
the central charge. If we shift the edge by a small amount $\delta
A$,
this term of order $1/L$ will have a variation  of order
$1/L^2$  which we do not look for. Only the variation of the
extensive term
is therefore of interest. We have
$$
{\tilde{E}\over L}=\int_{-\infty}^{A+\delta A} \left(\mu
e^\theta-{V\over
2}\right)\tilde{\rho}(\theta)d\theta
$$
where
$$
\tilde{\rho}(\theta)={\mu e^\theta\over 2\pi}+
\int_{-\infty}^{A+\delta
A}
\Phi(\theta-\theta')\tilde{\rho}(\theta')d\theta'.
$$
As in \korbook\ we can easily rewrite
$$
{\tilde{E}\over L}={\mu \over 2\pi}\int_{-\infty}^{A+\delta A}
\tilde{\epsilon}(\theta)e^\theta d\theta
$$
and  a similar equation for non tilde quantities,
where
$$
\tilde{\epsilon}(\theta)=\mu e^\theta-{V\over
2}+\int_{-\infty}^{A+\delta
A}
\Phi(\theta-\theta')\tilde{\epsilon}(\theta')d\theta'.
$$
To get a corection of order $1/L$ to the energy, we expand
$\tilde{E}$
and $\tilde{\epsilon}$  as functions of $\delta A$. One finds
$$
\eqalign{
\left.{\partial\tilde{\epsilon}(\theta)\over\partial \delta
A}\right|_
{\delta A=0}&=0\cr
\left.{\partial^2\tilde{\epsilon}(\theta)\over\partial^2\delta A}
\right|_
{\delta A=0}&=\dot{\epsilon}(A)L(\theta,A),\cr}
$$
from which it follows that the first derivative of $E$ at $\delta
A=0$ vanishes,
and for the second derivative one has
$$
{1\over L}\left.{\partial^2\tilde{E}\over\partial^2\delta A}
\right|_
{\delta A=0}={\mu\over 2\pi}\dot{\epsilon}(A)\left[{V\over 2}+{V\over
2}\int_{-\infty}^A
L(\theta,A)d\theta\right]
$$
It follows finally that
\eqn\maincft{\tilde{E}-E= \mu L {gV^2\over 8\pi}(\delta A)^2.}
Now suppose we want to create an excitation of charge $Q$. For each
volton added,
a certain number of voltons leave the Fermi sea, so for $dn$ voltons
added, we have a charge $Q=\left[1+F_b(-\infty|A)\right]dn$.
By which amount does the Fermi edge shift if we add
one volton with minimum possible energy? First, the rapidity
of the soliton immediately below the added volton shifts,
by an amount $\delta_1={F(A|A)\over L\rho(A)}$. Second,
the added volton must lie immediately above, at a rapidity
differing from this one by $\delta_2={1\over L\rho(A)}$. This leads
to $\delta A={F(A|A)\over L\rho(A)}+{1\over L\rho(A)}$
and thus a change of energy, using \maincft\
$$
\tilde{E}-E={1\over L}{gV^2\over 8\pi}\left({F_b(A|A)+1\over
\rho(A)}\right)^2
$$
If we add $dn$ voltons, the change of energy will thus read
$$
\tilde{E}-E={\pi\over L}[1+F_b(A|A)]^2(dn)^2.
$$
Using the formulas for the charge, together with the
usual formula between the gaps and the conformal weights
\ref\C{J. Cardy, J. Phys. A17 (1984) L385} leads to
\eqn\maincftii{h={1\over 2}[1+F_b(A|A)]^2={1\over
4g}[1+F_b(-\infty|A)]^2(dn)^2={Q^2\over 4g}.}
This agrees with what is expected from the Lagrangian. Notice that,
once
again,
the result holds for any value of $\delta_b$: so far, we have
obtained no
constraint for this unknown parameter. Its value is however
crucial for the stattering theory, as we now discuss.

\subsec{Renormalized S matrix}

To proceed, we consider the S matrix of excitations. Take
for instance  two voltons at rapidities $\theta_1,\theta_2$,
with $\theta_1>\theta_2$.
With the first volton only, the rapidities in the sea are shifted
to $\theta_\alpha^{(1)}$, with both particles they
are shifted to $\theta_\alpha^{(12)}$. Passing the first volton
through the system in the presence of the (shifted) Fermi sea
only, one gets a phase $\phi_1$, while passing it
through the system in the presence of the (shifted) Fermi sea and the
second  particle,
one gets a phase $\phi_{12}$. These two phases read respectively,
\eqn\phasexpres{\eqalign{\phi_1=&L e^{\theta_1}+2\pi \sum_\alpha
\delta(\theta_1-
\theta_\alpha^{(1)})\cr
\phi_{12}= &L
e^{\theta_1}+2\pi\sum_\alpha\delta(\theta_1^p-\theta_\alpha^{(12)})+
\delta(\theta_1-\theta_2).\cr}}
Straightforward computations give
\eqn\sdef{\phi_{12}-\phi_1=-2\pi F_b(\theta_1|\theta_2),}
wiht $F$ given in \explic. Here, we have assumed that
with $n$ voltons, the boundary conditions will be  changed so that
a term $n\delta_b$ will be added to the left
hand side of quantization equations (similar to \condsi)
for solitons in the sea.  We then define the S-matrix of voltons by
\eqn\svolt{
S_{++}(\theta_1,\theta_2)=-\exp\left[-2i\pi F_b(\theta_1|\theta_2)
\right],\ \theta_1>\theta_2.}
The previous computation does not require $\theta_1>\theta_2$ to
be algebraicaly correct. What does require  this inequality
however is the identification of the $S$ matrix with the
physical process of passing the particle 1 through the system.

We now build a Faddeev Zamolodchikov \ZZ\ algebra for our
excitations,
setting
\eqn\FZi{
|\theta_1,\theta_2>_{++}=S_{++}
(\theta_1,\theta_2)|\theta_2,\theta_1>_{++},
\ \theta_1>\theta_2.}
{}From this, we see that
\eqn\FZii{
|\theta_1,\theta_2>_{++}=S_{++}^{-1}(\theta_2,\theta_1)
|\theta_2,\theta_1>_{++},
\ \theta_1<\theta_2.}
We will henceforth set
\eqn\defS{\eqalign{S_{++}(\theta_1,\theta_2)=-\exp\left[-2i\pi
F_b(\theta_1|\theta_2)
\right],\ \theta_1>\theta_2\cr
=-\exp\left[2i\pi F_b(\theta_2|\theta_1)
\right],\ \theta_2>\theta_1.\cr}}
The S-matrix satisfies
\eqn\uniti{S_{++}(\theta_1,\theta_2)S_{++}(\theta_2,\theta_1)=1.}
It is  singular at $\theta_1=\theta_2$, and is not a function
of $\theta_1-\theta_2$ only.
The same analysis for antivoltons results in
\eqn\Sh{\eqalign{S_{--}(\theta_1,\theta_2)=&-
\exp\left[-2i\pi F_b(\theta_1|\theta_2)\right],\ \theta_1<\theta_2\cr
&-
\exp\left[2i\pi F_b(\theta_1|\theta_2)\right],\
\theta_1>\theta_2.\cr}}
Observe that the phases are opposite to the
ones in \defS. This is because the momentum of an antivolton,
$p=p_f+\epsilon(\theta)$, is a decreasing function of $\theta$ for
$\theta<A$.
Finally, one can also scatter a
volton  and an antivolton. One has
\eqn\Sphdef{|\theta_1,\theta_2>_{+-}=S_{+-}(\theta_1,\theta_2)
|\theta_2,\theta_1>_{-+},}
with
\eqn\Sph{S_{+-}(\theta_1,\theta_2)=-\exp\left[2i\pi
F_b(\theta_1|\theta_2)\right],}
together with
\eqn\Shp{S_{-+}(\theta_1,\theta_2)=-\exp\left[-2i\pi
F_b(\theta_2|\theta_1)\right].}
In the foregoing equations, there
is an additional minus sign compared with the phase shifts like
\sdef\  because
our definition of the S-matrix is through  the Faddeev Zamolodchikov
algebra
\ref\KM{T. Klassen, E. Melzer, Int. J. Mod. Phys. A8 (1993) 4131.}.
Indeed, consider for instance a pair of particles in some scattering
theory. If
$\phi_{12}$ is the phase one gets passing the first through the
second (this
$\phi_{12}$ should not be confused with the one in equation \sdef),
the coordinate wave function reads, before symmetrization or
antisymmetrization
$$
\eqalign{e^{-i\phi_{12}/2}e^{i(p_1x_1+p_2x_2)},\ x_1<x_2\cr
e^{i\phi_{12}/2}e^{i(p_1x_1+p_2x_2)},\ x_1>x_2\cr}
$$
from which
$$
\eqalign{|p_1p_2>=&\int_{x_1<x_2}e^{-i\phi_{12}/2}
e^{i(p_1x_1+p_2x_2)}
\chi^+(x_1)\chi^+(x_2)|0>\cr
+&\int_{x_1>x_2}e^{i\phi_{12}/2}e^{i(p_1x_1+p_2x_2)}
\chi^+(x_1)\chi^+(x_2)|0>\cr
=&-e^{-i\phi_{12}}|p_2p_1>,\cr}
$$
where we used that the $\chi$ operators are
 fermionic, and $\phi_{12}+\phi_{21}=0$.

\subsec{Form-factors}

Still assuming a generic $\delta_b$, let
us consider the physics at low energy. Right moving, low energy
excitations are made
of voltons antivoltons pairs near the Fermi rapidity $\theta=A$.
Because
there is an energy scale $V$, the theory is not relativistically
invariant, even though neutral excitations have dispersion relation
$\epsilon=p$. This means, as observed above, that the S-matrix does
not depend only on the ratio of energies. However, very close to $A$,
the ratio $\epsilon/V$ goes to zero while the S-matrix
elements go to constants.  In that limit, we deal again with
a relativistically invariant theory, now depending on
$\delta_b$.
The key parameter now is
\eqn\key{\kappa\equiv F_b(A|A)=\sqrt{2g}
[\delta_b-\delta(\infty)]+\sqrt{2g}-1.}
Since the theory is relativistic, it is convenient to
parametrise the energy in terms of a renormalised rapidity
$\beta$ (see I).
For voltons, we set $\epsilon(\theta)=\mu e^{\beta},\theta>A$,
and for
antivoltons, we set $\epsilon(\theta)=-\mu e^{\beta},\theta<A$.
We normalize  asymptotic states
to $2\pi\delta(\beta_1-\beta_2)$. Note that due to the minus sign,
for antivoltons,
the function $\beta(\theta)$ is decreasing so
increasing momentum means increasing renormalized rapidities.

All form factors are now expected to depend on differences of
$\beta$'s,
which are renormalized rapidities. The choise $b=b^*$ or
$\kappa=0$ is the
only one for which the S-matrix is analytic at coincident rapidities.
This by itself is probably enough to dictate that choice. However,
we wish to accumulate more evidence, and keep studying
the generic case first.

We discuss  the form factor $<0|j(z)|\beta_2,\beta_1>_{-+}$,
where the right hand side physically describes
a hole particle pair and $z=x-t$. The $z$ dependence
of this correlator
follows trivially from kinematic considerations, $\exp
\mu iz[e^{\beta_1}+e^{\beta_2}]$, so we restrict to the case $z=0$.
We
call  this form factor $f_{+-}(\beta_1,\beta_2)$\foot{The notation
is similar to \smirnov,
$<0|j|\beta_2,\beta_1>=f(\beta_1,\beta_2)$.}.
The first axiom to be satisfied is
\eqn\axiomi{f_{+-}(\beta_1,\beta_2)S_{+-}(\beta_1,\beta_2)
=f_{-+}(\beta_2,\beta_1)}
with the $S$ matrix given by its low energy limit described
by \key .
An obvious solution to \axiomi\ is simply to set
\eqn\axiomii{\eqalign{f_{+-}(\beta_1,\beta_2)\propto
\exp[-i\pi\kappa]\cr
f_{-+}(\beta_2,\beta_1)\propto-\exp[i\pi\kappa] .\cr}}
The second axiom is
\eqn\axiomiii{\eqalign{
f_{+-}(\beta_1,\beta_2+2i\pi)&
=S_{+-}(\beta_1,\beta_2)f_{+-}(\beta_1,\beta_2)\cr
f_{+-}(\beta_1+2i\pi,\beta_2)=&S_{+-}^{-1}(\beta_1,\beta_2)
f_{+-}(\beta_1,\beta_2).\cr}}
And a solution to \axiomiii\  is of the form
\eqn\soli{f_{+-}[\beta_1,\beta_2]\propto
\mu e^{\beta_1/2}e^{\beta_2/2}\left(e^{\beta_2-\beta_1}\right)^\kappa
,}
where we have put
the correct  dimension  $1/length$ of the form factor. Apart from the
dimensional factor, the form factor is expected to depend on the
massless relativistic invariant $s\equiv e^{\beta_2-\beta_1}$. We see
from
\soli\ that there is a cut in the $s$ plane along the negative real
axis.
With $\kappa\neq 0$, such cut is unavoidable since the S matrix
is singular at coincident rapidities.

The last axiom in the relativistic case stems from crossing:
\eqn\axiomiv{_-<\beta_1|j(0)|\beta_2>_-=
f_{+-}(\beta_1-i\pi,\beta_2).}
Leading
to the minimal conjecture for the two particle form-factor
\eqn\minim{
f_{+-}(\beta_1,\beta_2)=i\mu
q_-e^{-i\pi\kappa}e^{\beta_1/2}e^{\beta_2/2}
\left(e^{\beta_2-\beta_1}\right)^\kappa.}
The normalisation of the form factor is related
to the charge term which originates from crossing,
as we now demonstrate. Indeed, we must satisfy the relation
\eqn\chargenorm{\int_{-\infty}^\infty\  _-<\beta_1|j(z)|\beta_2>_-
dz=
 2\pi\delta(\beta_1-\beta_2) q_-.}
After crossing, performing the integral over $z$,
constrains $\beta_1=\beta_2$ and brings a jacobian. The ratio of
$\epsilon$'s
becomes equal to unity and disappears. The only remaining
part, therefore is the charge $q_-$, as desired.
In the case $\kappa=0$,  $\delta_b=\delta_{b^*}=-{1\over 2}+
{1\over \sqrt{2g}}-{1\over 4g}$,
$q_-=-\sqrt{2g}$, so one has then
\eqn\minim{f_{+-}(\beta_1,\beta_2)=
-i\mu\sqrt{2g}e^{\beta_1/2}e^{\beta_2/2}.}
Note that instead of \axiomii, \axiomi\ could be solved also
by setting,
\eqn\tatay{
\eqalign{f_{+-}(\beta_1,\beta_2)\propto
\left(e^{\beta_1}-e^{\beta_2}\right)^
\kappa
\cr
f_{-+}(\beta_2,\beta_1)\propto
-\left(e^{\beta_2}-e^{\beta_1}\right)^\kappa.\cr}}
However, in the computation of the
charge, this would result in an additional factor $e^{\kappa\beta}$
in the right hand side of \chargenorm, which is impossible.

The same logic can be applied to form-factors
with higher number of particles. A reasonable guess for the four
particle form-factor
is
\eqn\fourp{\eqalign{
<0|j(z)|\beta_4,\beta_3,&\beta_2,\beta_1>_{--++}\propto
q_-\mu(e^{2i\pi\kappa}-1)\left(e^{\beta_3+\beta_4-\beta_1-\beta_2}
\right)^\kappa
\left(e^{\beta_1}+e^{\beta_2}
+e^{\beta_3}+e^{\beta_4}\right)\cr
&{\sinh{\beta_1-\beta_2\over 2}\sinh{\beta_3-\beta_4\over 2}\over
\cosh{\beta_1-\beta_3\over 2}\cosh{\beta_1-\beta_4\over 2}
\cosh{\beta_2-\beta_3\over 2}\cosh{\beta_2-\beta_4\over 2}},\
\beta_1>\beta_2, \
\beta_3>\beta_4.\cr}}
Form factors where the $\beta$'s  have been
exchanged will take similar forms, up to phases arising from the
S-matrix
as in \axiomii\ -
reproducing these phases
by terms depending on the differences of the energies would
make the pole axioms and the
charge computation impossible as in \tatay. Note
that these phases are singular   at coincident rapidities
$\beta_1=\beta_2$ or
$\beta_3=\beta_4$,  since the $S$ matrix is.

 Expression \fourp\
vanishes when two voltons or antivoltons coincide, as physically
required in the Bethe ansatz.
It also has a simple pole as a volton and
antivolton annihilate, eg $\beta_4=\beta_2+i\pi$, with a residue
that is proportional to the two particle form-factor in agreement
with the  axioms
in \smirnov\ (it is not completely
clear how exactly to generalize the pole axiom,
and therefore what this residue exactly should be).

Let us now use these form factors to discuss physical
quantities, relying heavily on I for the definitions
of the impurity problems and the quantities under study.
The important point about \fourp\ is that it transforms as $FF\to
e^{\sigma}FF$
when $\beta\to\beta+\sigma$. As a result it will, together with
all the higher form factors, contribute to the noise computation
at order $\omega$. While the normalisation of successive form-factors
are dictated by LSZ reduction formulas \smirnov, and the
normalisation of the first
form-factor by the charge,  a very non trivial
sum rule has to be satisfied to ensure that the current correlator
reproduces
\eqn\noisedef{S(\omega)=\int e^{i\omega t}\left<\{j(t),j(0)\}\right>=
{g|\omega|\over\pi},}
(see I).
We suspect
this sum rule would not be satisfied away from $\kappa=0$,
but we have no proof of this.

Some restrictions are obvious however. The two particle form factor
by itself gives to the noise a contribution (assuming $\omega>0$)
\eqn\usefeqi{
S^{(2)}= {1\over 2\pi}(q_-)^2 \int_0^\omega dx \left({x\over
\omega-x}\right)^{2\kappa}
={2\kappa\pi \over\sin 2\kappa\pi}(1/2+g+2g\delta_b)^2 {\omega\over
2\pi}.}
The ratio involving $\kappa$ is clearly greater than one. Hence one
needs $(1/2+g+2g\delta_b)^2\leq 2g$.
For instance, for $g<1/2$, this excludes the choice $\delta_b=0$. The
choice
$\delta_b=1/2$ is always excluded. In fact, any choice $\delta_b=cst$
is excluded
for $g$ small enough, so $\delta_b$ must have some non trivial
dependence on $g$.

\subsec{$\chi''$ revisited.}

Following I, let us now add a boundary interaction. Without a
voltage,
the effect of the boundary is taken into account by  reflection
matrices $R$  (solutions of the boundary Yang-Baxter equation)
for the ``bare'' excitations. With the
voltage, one obtains new, dressed reflection matrices ${\cal R}$, for
the excitations over the Fermi sea.

Consider again the computation
of the low frequency behaviour of $\chi''(\omega)$
in dissipative quantum mechanics along the same  lines
as in I. For a generic value of $\kappa$,
the two particle term would be proportional to, instead of eq. (5.8)
in I,
\eqn\compi{\int_0^\omega [{\cal R}(x){\cal
R}^*(x-\omega)-1]\left({x\over \omega-x}\right)^{2\kappa} dx.}
Change variables $x\to \omega x$ to get
\eqn\compii{\omega\int_0^1 [{\cal R}(\omega x){\cal
R}^*(x\omega-\omega)-1]\left({x\over 1-x}\right)^{2\kappa} dx.}
We are interested in the first non trivial real term, which
occurs from the expansion of the integrand to order $\omega^2$.
Setting ${\cal R}=e^{i\phi}$, the bracket gives a total of four terms
$$
-{1\over 2}(\phi')^2\omega^2[x^2+(x-1)^2-2x(x-1)]=-
{1\over 2}(\phi')^2\omega^2
$$
so the first contribution is equal to
$$
-{1\over 2}(\phi')^2\omega^3\int_0^1 \left({x\over
1-x}\right)^{2\kappa} dx
$$
Observe that this integral is exactly the one that would appear
in the computation of the current correlator without impurity
\usefeqi: the impurity dependence ($\phi'$)
and the energy dependence actually factor out.  This result is
true at every order. For instance  the next order term would we
proportional to
$$
\eqalign{
&\int_0^\omega\int_0^\omega\int_0^\omega|f(x_1,x_2,x_3,
\omega-x_1-x_2-x_3|^2\cr
&[{\cal R}(x_1){\cal R}(x_2){\cal R}^*(-x_3){\cal
R}^*(x_1+x_2+x_3-\omega)-1]
{dx_1dx_2dx_3\over x_1x_2x_3(\omega-x_1-x_2-x_3)},\cr}
$$
where $f$ is the four particle form-factor \fourp. Now the same
integral
without the term in brackets is the one appearing
in the computation of the noise, the fourth order term of a whole
series that sums up to $g\omega$. Contributing to
the linear term of $\chi''$ we have to extract
the term of third order in $\omega$. Make a change of variables
$x\to \omega x$. Using homogeneity of the
current form-factors, we can rewrite the integral as
$$
\eqalign{
&\omega\int_0^1\int_0^1\int_0^1|f(x_1,x_2,x_3,1-x_1-x_2-x_3|^2\cr
&\{{\cal R}(\omega x_1){\cal R}(\omega x_2){\cal R}^*(-\omega x_3)
{\cal R}^*[\omega(x_1+x_2+x_3-1]-1\}
{dx_1dx_2dx_3\over x_1x_2x_3(1-x_1-x_2-x_3)}\cr}
$$
To get $\chi''$, we expand the integrand
to order $\omega^2$. Setting again  ${\cal R}=e^{i\phi}$, the
corresponding terms
are identified,
 setting $x_4=1-x_1-x_2-x_3$, as
$$
-{1\over 2}(\phi')^2\omega^2\left[\sum
x_i^2-2\sum_{i<j}x_ix_j\right]
=-{1\over 2}(\phi')^2\omega^2
$$
Hence the term $-{1\over 2}(\phi')^2\omega^2$
factors out, and the remaining integral
is again {\bf the same} as the one arising in the
computation of the correlator without
impurity.  Since the form-factors are
normalized to sum up these integrals to  the same term $g\omega/\pi$
for
 any $\delta_b$, we get the
same formula  for  the limit (as $\omega\to 0$) of
$\chi''(\omega)/\omega$ for any value of $\delta_b$!
Observing that $\phi'={\alpha\over i}e^{-\theta}\left.
{d\over d\theta}\ln {\cal R}(\theta)\right|_{\theta=A}$, where
$\alpha={d e^\theta\over d\epsilon(\theta)}$
is given in I,  we recover
\eqn\sameoldstuff{\lim_{\omega\rightarrow 0}{\chi''(\omega)\over
\omega}=-{\alpha^2\over 4g\pi^2}
\left(e^{-\theta}\left.{d\over d\theta}\ln {\cal
R}(\theta)\right|_{\theta=A}\right)^2.}
This is true for any value of $\delta_b$.

\subsec{Summary}

The analysis of the theory of excitations over the new ground
state, when done carefully, turns out to involve an extra phase
$\delta_b$, which has a key influence  on the dressed
scattering theory, and thus on the response functions
in the presence of a voltage.  In this long
analysis we have looked at
various physical quantities to check whether we could fix
$\delta_b$.
The computation of $\lim_{\omega\rightarrow 0}{\chi''(\omega)\over
\omega}$ and the corresponding Shiba relation of I are not enough
to settle the value of $\delta_b$: in fact, any value would give the
right
result, although the value that leads to $\kappa=0$ is the
most reasonable physically, and actually the only one for which
the sum rules for the charge and noise can be explicitely checked.
In contrast with dissipative quantum mechanics, the value of $\kappa$
is crucial
 for tunneling between edges in  quantum Hall devices, also discussed
in I.
Indeed, if $\kappa$ was non zero and the theory admitted multiple
particle, low energy excitations,
the noise in the presence of an impurity
would involve  terms of an arbitrarily high degree
in the reflection matrix. The same feature would be observed
in the conductance, as will be discussed later in this paper.
 This is in contradiction
with formulas that have been derived using a variety of approaches
(in particular
a Boltzmann type equation \flsbig, or a
 Landauer-B\"uttiker approach \LB, \FSnoise), and checked against
duality,
perturbative expansions, numerical and experimental data. To
summarize: while we
have no proof that $\kappa=0$, there is a large body of practical
and conceptual evidence that it is so.

\newsec{Beyond the low energy behaviour}

Let us thus  assume from now on that $\delta_b=\delta_{b^*}$
 has been chosen so that
$\kappa=0$,
and that,  at low energy, the theory
is made of free fermions. What can we say for higher
energies? Let us restrict to energies $\omega<gV$ so
the only processes involved concern volton-antivolton
pairs. Since excitations have finite
energy, we turn back to the original variables $\theta$. We consider
the form-factor $<0|j(z)|\theta_2,\theta_1>_{-+}$. The $z$
dependence is now $\exp iz[\epsilon(\theta_1)-\epsilon(\theta_2)]$.
Since
the S-matrix describes exchange properties, it is natural to expect
that
the analog of \axiomi\ will be satisfied (we denote $F_{b^*}\equiv
F$),
\eqn\newaxiomi{f_{+-}(\theta_1,\theta_2)S_{+-}(\theta_1,\theta_2)
= f_{-+}(\theta_2,\theta_1).}
Generalizing \axiomii, this has the simple
solution
\eqn\newaxiomii{\eqalign
{f_{+-}(\theta_1,\theta_2)\propto \exp[-i\pi F(\theta_1,\theta_2)]\cr
f_{-+}(\theta_2,\theta_1)\propto -\exp[i\pi
F(\theta_1,\theta_2)].\cr}}
The analog of crossing is also clear. While voltons
are originally defined for $\theta>A$ and antivoltons for $\theta<A$,
one can continue the associated amplitudes
by analyticity through the threshold $A$. Creating an antivolton
at $\theta>A$ will simply mean, as before, destroying a soliton
(called
a volton here) at this rapidity. In other words, we expect
the analog of \axiomiv:
\eqn\newaxiomiv{_-<\theta_1|j(0)|\theta_2>_-
=<0|j(0)|\theta_2,\theta_1>_{-+}=f_{+-}(\theta_1,\theta_2).}
The shift of $i\pi$ was necessary in the $\beta$
parametrization of the relativistic case
to describe both sides of the thresholds, something
we simply accomplish here by continuation through the
threshold.

Away from the relativistic limit, it is not clear
what the equivalent of  \axiomii\ should be. However, the charge
normalization can still be used. Indeed,
even for excitations of finite energy,
since $F(-\infty|\theta_p)$ is independent of
$\theta_p$, the renormalized charge is as before $\sqrt{2g}$. From
\newaxiomiv\
we conjecture
\eqn\newminim{f_{+-}(\theta_1,\theta_2)=
-i\sqrt{2g}\exp[-i\pi F(\theta_1|\theta_2)][\dot{\epsilon}
(\theta_1)]^{1/2}
[\dot{\epsilon}(\theta_2)]^{1/2}
\left(-{\epsilon(\theta_2)\over
\epsilon(\theta_1)}\right)^{F(\theta_1|\theta_2)}.}
Here, $\dot{\epsilon}\equiv{d\epsilon\over d\theta}$. This formula
agrees with \minim\ in the relativistic limit  once
the change of normalization
$\delta(\theta_1-\theta_2)\to\delta(\beta_1-\beta_2)$
is taken into account. The last term is fixed by dimensional
analysis,
together with the charge normalization (it is necessary to off set
the overall $\exp[-i\pi F(\theta,\theta)]$  at coincident
rapidities).

It is interesting to discuss  this form factor by getting back to the
noise without
impurity. The contribution of the two particle form
factor is proportional to  (assuming $\omega>0$)
\eqn\tutui{
\int_A^\infty\int_A^\infty
|f_{+-}(\theta_1,\theta_2)|^2\delta[\epsilon(\theta_1)-
\epsilon(\theta_2)-\omega]d\theta_1d\theta_2=
\int_{A}^{\epsilon^{-1}(\omega)} {1\over
\dot{\epsilon}(\theta)}|f_{+-}(\theta, \epsilon^{-1}
[\epsilon(\theta)-\omega])|^2d\theta.}
At first order, we simply set $q=\sqrt{2g}$, $\theta=A$,
to get
$$
(\sqrt{2g})^2 \int_{A}^{\epsilon^{-1}(\omega)} \dot{\epsilon}(\theta)
d\theta=2g\omega
$$
as requested. At next order, we have to take the other terms into
account.
 Setting $\theta_1=\theta$, observe
that
$$
\eqalign{F(\theta_1|\theta_2)=&L(A,A)(\theta_1-\theta_2)\cr
=&L(A,A)(\theta_1-A+A-
\theta_2)\cr
=&{L(A,A)\over \dot{\epsilon}(A)}[\epsilon(\theta_1)-
\epsilon(\theta_2)]\cr
=&\lambda\omega\cr}
$$
Here we defined
\eqn\lamdef{\lambda\equiv {L(A,A)\over \dot{\epsilon}(A)}=
{L(A,A)\over 2\pi\rho(A)}.}
The contribution of this term to the noise reads
$$
\eqalign{
(\sqrt{2g})^2 \int_A^{\epsilon^{-1}(\omega)} \dot{\epsilon}(\theta)
d\theta \left( {\epsilon(\theta)\over \omega-\epsilon(\theta)}\right)
^{\lambda\omega}\cr
=2g\omega\int_0^1 \exp\left[ \lambda\omega\ln {x\over 1-x}\right]
\cr}
$$
Expanding the exponential to first order gives no $\omega^2$
correction
due to $\int_0^1 \ln {x\over 1-x}=0$.

Of course, the form factor contributes a non vanishing term at order
$\omega^3$.
This should be offset by the next  form-factor
involving a pair of voltons and a pair of antivoltons. Indeed, while
in
the relativistic limit with $\kappa\neq 0$, such a form factor
gave a contribution of order $\omega$ to the noise, now
the residue axiom, reasonably generalized to the
non-relativistic case, will lead to a form factor similar to \fourp,
but with the term $e^{2i\pi\kappa}-1$ replaced
by a term of order ${\omega\over V}$, measuring the difference
between the S-matrix and $-1$ away from the Fermi rapidity. In the
computation
of the noise, this will give a term of order $\omega (\omega/V)^2$.
More
generally, the process involving n voltons and n antivoltons
will give a leading contribution of order $\omega (\omega/V)^{2n-2}$.

Let us now discuss in more  details the term $\omega^3$. The integral
 \tutui\
gives for the noise a contribution (assuming $\omega>0$)
\eqn\tutuii{S(\omega)^{(2)}={g\omega\over\pi}+\zeta \omega^3,}
where  the value of $\zeta$ will not be needed in what
follows.
Since  the total noise is $S(\omega)={g\omega\over\pi}$, this  means
that the 4 particle form factor
must contribute to the noise by a leading term
\eqn\tutuiv{S(\omega)^{(4)}=-\zeta\omega^3,}
and thus we obtain the sum rule for the four particle form factor
\eqn\tutv{\int_A^\infty
|f_{++--}(\theta_1,\theta_2,\theta_3,\theta_4)|^2
2\pi\delta[\epsilon(\theta_1)+\epsilon(\theta_2)-\epsilon(\theta_3)-
\epsilon(\theta_4)-\omega] \prod_{i=1}^4 {d\theta_i\over
2\pi}=-\zeta\omega^3.}

The point is, that this sum rule is enough to determine the next term
in $\chi''(\omega)$. Indeed, the two particle form factor
contributes to $\chi''$ by a term proportional to
\eqn\tutuvi{\int_A^\infty\int_A^\infty
|f_{+-}(\theta_1,\theta_2)|^2\delta[\epsilon(\theta_1)-
\epsilon(\theta_2)-\omega]\hbox{Re} [{\cal R}(\theta_1){\cal
R}^*(\theta_2)-1]
d\theta_1d\theta_2.}
We are now interested in getting the terms of order $\omega^3$ and
$\omega^5$
in this integral. The term of order $\omega^3$ was already evaluated
in I.
For the term of order $\omega^5$, since the integral combined
with the delta function
contributes an overall $1/\omega$, there will be two
contributions: we can either expand the ${\cal R}$
bracket to order $\omega^4$ and take the leading expression
($O(\omega^2)$)
 for the two
particle form-factor, or expand the  ${\cal R}$  bracket only to
order $\omega^2$ and take the next to leading
contribution for the two particle form factor ($O(\omega^4)$).

Similarly, the four particle form factor contributes by a term
proportional to
\eqn\tutuvii{\eqalign{\int_A^\infty
|f_{++--}(\theta_1,\theta_2,\theta_3,\theta_4)|^2
\delta[\epsilon(\theta_1)+\epsilon(\theta_2)-\epsilon(\theta_3)-
\epsilon(\theta_4)-\omega]\cr
\hbox{Re} [{\cal R}(\theta_1)
{\cal R}(\theta_2){\cal R}^*(\theta_3){\cal R}^*(\theta_4)-1]
\prod_{i=1}^4 d\theta_i.\cr}}
Taking the term of order $\omega^6$ in the four particle form factor
would necessitate taking the  order 0 in the ${\cal R}$  bracket,
which vanishes. Hence, the only term we need
to consider is  order $\omega^2$ in the ${\cal R}$ bracket, and the
leading order  ($O(\omega^4)$) in the four particle form factor.  Now
the point is,
that like in the previous
section (paragraph $2.5$), the term of order  $\omega^2$  in the
${\cal R}$ bracket is actually independent
of the rapidities, and simply factors out as a constant. The integral
that is left
is the same as the one appearing in the noise. The same remark holds
for the two particle form-factor
contribution \tutuvi, so these two terms cancel out! Hence, all what
remains to
determine the $\omega^5$ order is the  ${\cal R}$  bracket
to order $\omega^4$, combined with the leading expression
($O(\omega^2)$)
 for the two
particle
form-factor. In other words, the expression that was obtained in
 I is good to get the
$\omega^5$
order:
\eqn\gri{\chi''(\omega)=-{1\over 2g\pi^2\omega^2}\hbox{Re }
\int_{-\infty}^{\ln\omega}d\beta_2 d\beta'_2\left[ {\cal
R}^*(\theta_2){\cal
R}(\theta'_2)-1\right]e^{\beta_2}e^{\beta'_2}
\delta(\omega-e^{\beta_2}-e^{\beta'_2}),}
Here we have reparametrized $\epsilon\to e^\beta$, which defines the
$\theta\to\beta$ correspondence.
Redefining ${\cal R}(\epsilon)={\cal R}(\theta)$ we rewrite \gri\ as
\eqn\gribis{\eqalign{
\chi''(\omega)&=-{1\over 2g\pi^2\omega^2}\hbox{Re
}\int_0^\omega dx
[{\cal R}(x){\cal R}^*(x-\omega)-1]dx\cr &
=-{1\over 2g\pi^2\omega}\hbox{Re
}\int_0^1
dx[{\cal R}(\omega x){\cal R}^*(x\omega-\omega)-1].}}
Laborious but straightforward manipulations lead to the expression
\eqn\gribisi{{\chi''(\omega)\over \omega}\approx
{(\phi')^2\over 4g\pi^2}+{1\over 2g\pi^2}
\left[ {(\phi'')^2\over 24}+{\phi'\phi'''\over 15}-
{(\phi')^4\over 40}\right]\omega^2+O(\omega^4)\ ,\ \omega\to 0}
where we have set as usual ${\cal R}=e^{i\phi}$, and primes denote
successive derivatives
with respect to the variable $x=\epsilon(\theta)$. Equation \gribisi\
can be made completely explicit
using the exact form of the dressed R matrix eq. (5.4) of I, together
with the relation between $\epsilon$ and $\theta$ as given
in eq. (2.13) of I. No simplification emerges,
and at this stage there seems little point in
giving a more explicit expression of the second term in \gribisi.

\newsec{The problem with $V=0,T>0$}

\subsec{Generalities}

When  $T\neq 0$, it is well known that thermal properties
can be computed  by evaluating correlators in a ``thermal ground
state'',
that is, any state that is characterized by the equilbrium densities
\korbook, \LLSS. For
the problem at hand, restricting once again to $g=1/t$,  recall
that the particles are soliton, antisoliton and the n-breathers
$n=1,2,\ldots, t-2$. Particles will be designated by the
generic label $j$.   Introduce the functions $\epsilon_j$ solution of
\eqn\tba{\epsilon_j(\theta)= \mu_je^\theta-T\sum_k
\int_{-\infty}^\infty
\Phi_{jk}(\theta-\theta')\ln
\left[1+e^{-\epsilon_k(\theta')/T}\right]d\theta'.}
Then, equilibrium densities are given by
\eqn\eqdens{\rho_j(\theta)={1\over 2\pi}{d\epsilon_j\over d\theta},}
and the filling fractions are
\eqn\fill{{\rho_j\over \rho^h_j}\equiv \xi_j={1\over
1+e^{\epsilon_j/T}}.}
The $\epsilon$'s diverge at infinity as in a free theory:
$\epsilon_j(\theta)\approx
\mu_je^\theta$, $e^\theta>>T$. At minus infinity, an interesting
consequence of the interaction is that the $\epsilon$'s
go to a constant of order $T$. Explicitely one has
\eqn\lmmm{e^{\epsilon_n/T}\approx (n+1)^2-1,
e^{\epsilon_{s,a}/T}\approx t-1,
\theta\to-\infty.}
It is well known \korbook\ (see also below) that the $\epsilon$'s are
excitation energies: destroying (creating) a particle of type $j$
decreases (increases) the energy by $\epsilon_j$. Hence, due to
\lmmm, there are no
single particle excitations of arbitrarily low energy:
the low energy excitations are, as in the voltage case,
fully obtained by particle hole pairs (there is a gap
of order $T$ for any processes that do not conserve the number
of particles). Contrarily to the
foregoing voltage case, such pairs can however be created at any
(neighbouring)
rapidities since at $T>0$, the filling fractions are less than one.

As in the voltage case, we would like to know the form factors
of the current. Observe that, since only particle hole pairs are
involved
in the very low energy limit, breathers won't appear (since the
current operator
changes the parity of the number of breathers)
and we can restrict to soliton/antsoliton excitations. To
address the form factors, we need as before the renormalized charge
and
S-matrix.

\subsec{Renormalized S matrix, charge and form-factors}

We start with the shift function. Suppose for instance
we add a particle of type $k$ at rapidity $\theta_p$. The shift
of particles of type $j$ obeys, similar to \shifti\
\eqn\newshf{F_{jk}(\theta|\theta_p)-\sum_l\int_{-\infty}^\infty
\Phi_{jl}(\theta-\theta')\xi_l(\theta')F_{lk}(\theta'|\theta_p)
=\delta_{b/jk}(\theta)-\delta_{jk}(\theta-\theta_p).}
Observe that, since we consider massless excitations around
arbitrary rapidities, we have allowed the additional phase shift
$\delta_b$ to depend upon $\theta$.

To proceed, let us assume there is only one type of excitations.
Dealing
with several types makes the notation more
cumbersome without changing the argument. Then the equation
for the shift reads
\eqn\smplei{F(\theta|\theta_p)-\int_{-\infty}^\infty
\Phi(\theta-\theta')
\xi(\theta')F(\theta'|\theta_p)=
\delta_b(\theta)-\delta(\theta-\theta_p).}
The excitation energy for adding this particle is
\eqn\smpleii{\varepsilon(\theta_p)=\mu
e^{\theta_p}+\int_{-\infty}^\infty
\mu e^{\theta'}\xi(\theta')F(\theta'|\theta_p).}
Note that, compared with \dirI, there is
no term arising from $-\infty$ since without
a voltage, the bare (kinetic) energy of particles vanishes
in that limit. From \smpleii\ we deduce
\eqn\smpleiii{{d \varepsilon(\theta_p)\over d\theta_p}= \mu
e^{\theta_p}+
\int_{-\infty}^\infty
\mu e^{\theta'}\xi(\theta'){dF(\theta'|\theta_p)\over d\theta_p}.}
Now, from \smplei\ we have
$$
{dF(\theta|\theta_p)\over d\theta_p}-\int_{-\infty}^\infty
\Phi(\theta-\theta')
\xi(\theta'){dF(\theta'|\theta_p)
\over d\theta_p}=\Phi(\theta-\theta_p)
$$
Introduce the kernel, similar to \formal\
\eqn\newformal{{L(\theta|\theta')\over \xi(\theta')}=
\Phi(\theta-\theta')+\int_{-\infty}^\infty \Phi(\theta-\theta'')\xi(
\theta'')\Phi(\theta''-\theta')+\ldots.}
It follows that
$$
{dF(\theta|\theta_p)\over d\theta_p}={ L(\theta|\theta_p)\over
\xi(\theta_p)}
={L(\theta_p|\theta)\over \xi(\theta)}
$$
and thus
$$
{d \varepsilon(\theta_p)\over d\theta_p}=\mu e^{\theta_p}
+\int_{-\infty}^\infty
\mu e^{\theta'}L(\theta_p|\theta')d\theta'
$$
On the other hand, from \tba\ in the case of a single
particle, we deduce
\eqn\smpi{{d\epsilon\over d\theta}=\mu e^\theta+\int_{-\infty}^\infty
\Phi(\theta-\theta')\xi(\theta'){d\epsilon\over d\theta'}d\theta',}
so we find
$$
{d \varepsilon\over d\theta_p}={d\epsilon\over d\theta_p}
$$
hence $\varepsilon(\theta_p)=\epsilon(\theta_p)+\hbox{cst}$,
the constant depending on the function $\delta_b(\theta)$.

Consider now excitations around some rapidity $\theta_0$.
Shift the energy and momentum such that
these excitations have $e=p=\varepsilon(\theta)-
\varepsilon(\theta_0)
=\epsilon(\theta)- \epsilon(\theta_0)$. In the low energy limit,
parametrize
$e=e^\beta$, we are then in a situation similar to excitations
around the Fermi rapidity in the previous case. Chose the function
$\delta_b(\theta)$
such that $F(\theta_0|\theta_0)=0$ for any $\theta_0$. Then
the analog of $\kappa$ vanishes. To completely
determine the form factor of the current, we need
to determine the
charge of excitations. This is simpler
than in the case of finite voltage. Indeed, since solitons and
antisolitons
behave in the same way except for their charge, we chose the
same function $\delta_b$ for both species, and
$F_s(-\infty,\theta_p)=
F_a(-\infty,\theta_p)$. So, as many solitons as antisolitons
move across the rapidity $-\infty$, resulting
in a vanishing contribution to the charge.
In other words the renormalized charge equals the bare charge.
Going back to $\theta$ variables, at low energy $\theta_1\approx
\theta_2$,
\eqn\mank{<0|j(z)|\theta_2,\theta_1>_{-+}=-i[
\dot{\epsilon}_s(\theta_1)]^{1/2}[\dot{\epsilon}_s(\theta_2)]^{1/2}
e^{iz[\epsilon_s(\theta_1)-\epsilon_s(\theta_2)]},}
where we also used the symmetry (in the absence of applied voltage)
between solitons and antisolitons $\epsilon_s=\epsilon_a$.

\subsec{DC Conductance in the fractional quantum Hall effect using
Kubo formula}

{}From \mank, the $\omega\to 0$ limit of the noise without
impurity  follows
\eqn\nooset{\eqalign{\int dt &\left<\{j(t),j(0)\}\right>\cr
&=2\int_{-\infty}^\infty {d\theta_1d\theta_2\over 4\pi^2}
\xi[\epsilon_s(\theta_1)]\left(1-\xi[\epsilon_s(\theta_2)]\right)
\dot{\epsilon}_s(\theta_1)\dot{\epsilon}_s(\theta_2)2\pi
\delta[\epsilon_s(
\theta_1)
-\epsilon_s(\theta_2)-\omega]+(\omega\to -\omega)\cr
&= 2\int_{\epsilon_s(-\infty)}^\infty {dx\over 2\pi} {1\over
1+e^{-x/T}}
{1\over 1+e^{x+\omega/T}}
+(\omega\to -\omega)={4Tg\over 2\pi}+O(\omega^2).\cr}}
as required \foot{To compare with the usual form of Nyquist theorem,
recall that we have set $\hbar=1$, so the  dimensionless
conductance of the pure Luttinger liquid reads $G={g\over 2\pi}$.}.

In the presence of an impurity, we can also
compute the noise in a way similar to paper I. The LL and LR
noise take the same form. The RR noise is a bit different
because in the (thermal) ground state we have both
solitons and antisolitons this time. In fact the RR noise is easily
seen
to be the same as the LL noise here, as is easily proven by changing
basis $|\theta>^L_{\mp}\to P|\theta>^L_{\mp}+Q|\theta>^L_{\pm}$.
Hence
$$
\int dt \left<\{j(t),j(0)\}\right>={2\over 2\pi}\int_{-\infty}^\infty
\xi(\theta)
[1-\xi(\theta)]\dot{\epsilon}(\theta) |P(\theta)|^2d\theta.
$$
{}From this
we deduce the formula for the DC conductance
using the Kubo formula \foot{Here, we defined $G$ as $2\pi$ times
the usual conductance.}
\eqn\dccond{G={1\over T}\int_{-\infty}^\infty \xi_s(\theta)
[1-\xi_s(\theta)]\dot{\epsilon_s}(\theta) |P(\theta)|^2 d\theta,}
in agreement with \flsbig. Observe that, should multiple
particle excitations be allowed as discussed briefly in the voltage
case,
terms involving higher orders of $|P|^2$ would appear
in $G$. This seems unlikely in view of the success
of formula \dccond\ compared with numerical simulations. Also,
\dccond\ was initially derived using Landauer-B\"uttiker
scattering  formula, and although it is not entirely clear
whether one can apply it to the quasiparticles
of integrable systems, this alternate derivation
still increases our confidence in \dccond.

\subsec{Dynamical susceptibility in the double well problem}

As an application, we would like to consider
the low frequency behaviour of the  dynamical susceptibility
in the dissipative quantum mechanics problem and $T\neq 0$.
We need to introduce a dressed reflection matrix in that case,
which reads here
\eqn\dressedr{{\cal
R}_j(\theta)=R_j(\theta)\exp\left[\sum_k\int_{-\infty}^\infty
\xi_k(\theta')F_{kj}(\theta'|\theta){d\over d\theta'}\ln
R_k(\theta')\right].}
We define at $T\neq 0$,
\eqn\Xidef{\eqalign{\chi''(\omega)=&{1\over 2}\int {dt\over
2\pi}e^{i\omega t}
\left<[S^z(t),S^z(0)]\right>\cr
=&{1\over 2}\tanh {\omega\over 2T}\int {dt\over 2\pi}e^{i\omega t}
\left<\{S^z(t),S^z(0)\}\right>.\cr}}
The correlator of the spin anticommutator can be related with the
correlator of current operators.
The argument works exactly like at $T=0$ \LSS\ (the propagator
with Neumann boundary conditions
eq. (5.4) of \LSS\ becomes a rational function of trigonometric
functions
at finite $T$. When $x\to 0$, it vanishes everywhere
except at $y=y'$, and its integral for $y$ running
on the interval $[0,1/T]$ is still equal to 1).  Using  \LSS\ and
the previous form-factors, one finds at low frequency (
using again the symmetry between the two species at vanishing bias)
\eqn\dressedchi{{2T\over \omega}\chi''(\omega)\approx
-{1\over g^2\pi^2\omega^2}\hbox{Re }\int_{-\infty}^\infty
\xi[\epsilon_s(\theta)]\left(1-\xi[\epsilon_s(\theta)+\omega]\right)
\dot{\epsilon}_s(\theta)\left(
{\cal R}[\epsilon(\theta)]{\cal
R}^*[\epsilon(\theta)+\omega]-1\right)
d\theta
.}
Note that \dressedchi\ differs from \gri\ by factors of 2, due to the
fact
that excitations involve solitons or antisolitons, and then can have
positive
or negative energy. There is also a factor of $2g$ that differs,
due to the fact that the renormalized charge is $1$ with a
temperature
and no voltage, while it is $\sqrt{2g}$ with a voltage and no
temperature.
{}From \dressedchi\ one finds thus
\eqn\newdresschi{\lim_{\omega\to 0}{2T\over\omega}\chi''(\omega)=
-{1\over 2 g^2\pi^2}
\int_{-\infty}^\infty {\xi_s(\theta)[1-\xi_s(\theta)]\over
\dot{\epsilon}_s(\theta)}
\left({d\over d\theta}\ln {\cal R}\right)^2d\theta.}
In the case $g=1/2$, there is no dressing, and one finds
\eqn\freecase{\lim_{\omega\to 0}{2T\over\omega}\chi''(\omega)={8
\over \pi^2}
\int_0^\infty dx {1\over 1+e^{x/T}}{1\over 1+e^{-x/T}}
\left({T_B\over x^2+T_B^2}\right)^2,}
so, after using the transformation
$$
{1\over x+1}={1\over x-1}-{2\over x^2-1}
$$
one finds, using standard integral representations
of the $\psi$ function,
\eqn\gonehalfchi{\lim_{\omega\to
0}{2T\over\omega}\chi''(\omega)={1\over \pi^4 T}\left[{\pi T\over
T_B}\psi'\left({T_B\over 2\pi T}+{1\over 2}\right)
-{1\over 2}\psi''\left({T_B\over 2\pi T}+{1\over 2}\right)\right],}
in agreement with the result of \ref\SassWei{M. Sassetti, U. Weiss,
Phys. Rev. A41 (1990) 5383.}.

As small temperature, the sums are dominated
by particles whose rapidities approach $-\infty$. In that limit,
one checks that ${d\over d\theta}\ln {\cal R}\approx {4\pi\over
T_B}\rho$,
from which we recover $\chi''(\omega)\approx {\omega\over g(\pi
T_B)^2}$
at $T=0$, as shown in I.

Eq. \newdresschi\ can be made as explicit as necessary. Using the
relation
$$
{d\over d\theta}F_{kj}(\theta'|\theta)=
{L_{jk}(\theta|\theta')\over \xi_k(\theta')}
$$
one finds
$$
{d\over d\theta}\ln {\cal R}_j(\theta)={d\over d\theta}\ln
R_j(\theta)
+\sum_k\int_{-\infty}^\infty L_{kj}(\theta|\theta'){d\over d\theta'}
\ln R_k(\theta').
$$
Here, one has
$$
L_{jk}(\theta|\theta')=\Phi_{jk}(\theta-\theta')\xi_k(\theta')
+\int_{-\infty}^\infty \Phi_{jl}(\theta-\theta'')\xi_l(\theta'')
\Phi_{lk}(\theta''-\theta')\xi_k(\theta')+\ldots
$$
Since functions $\xi$ are easy to obtain by numerical solution
of the thermodynamic Bethe ansatz equations, it is
only a technical matter to determine \newdresschi.

\newsec{Conclusion}

We hope that the computations in this paper and
the previous one open a possible direction for the study of
correlators at finite $T$ and $V$ in quantum impurity
problems. Our results \gribisi,\newdresschi,
although completely explicit, are
unfortunately quite cumbersome: pushing the method further and
getting results  at arbitrarily large  frequency looks
much more involved than in the $T=0$ case, $V=0$ case (the
situation looks more favorable for problems which are massive
in the bulk, especially at low temperature \subir).
Nevertheless, interesting questions are already raised by this
approach: in particular,
the results of the present paper and
those
found in I do not agree with those of  perturbative methods. It would
be very useful to  investigate the matter further
in experiments or numerical studies such as
\ref\costi{T.A. Costi, C. Kieffer, Phys.
Rev. Lett. 76, 1683 (1996).}.

\vskip 0.4cm
\centerline{\bf Acknowledgements}

We thank C. de C. Chamon, S. Chakravarty,  D. Freed, A. Leclair and
S. Sachdev  for very useful discussions.

This work was supported by the Packard Foundation, the
National Young Investigator program (NSF-PHY-9357207) and
the DOE (DE-FG03-84ER40168).
F. Lesage is also  partly
supported by a canadian NSERC Postdoctoral Fellowship.

\listrefs

\end